\documentclass[12pt]{article}

\newcommand{\figurepath}{./}

\usepackage{setspace}
\usepackage{amsmath}
\usepackage{graphicx}
\usepackage{epsfig}
\usepackage{array}
\usepackage{rotating}
\usepackage{amsfonts}
\usepackage{ctable}
\usepackage{dcolumn}
\usepackage{multirow}
\usepackage{natbib}
\numberwithin{equation}{section}

\begin{document}
\title{Bull and Bear Markets During the COVID-19 Pandemic}

\author{John M. Maheu\footnote{DeGroote School of Business, McMaster
    University, maheujm@mcmaster.ca} \hspace{.1in} Thomas H. McCurdy\footnote{Rotman School of Management,
    University of Toronto, tmccurdy@rotman.utoronto.ca}
  \hspace{.1in}  Yong Song\footnote{Department of Economics, University of
    Melbourne, yong.song@unimelb.edu.au}}

\date{}
\date{November 2020}
\maketitle
\begin{abstract}
The COVID-19 pandemic has caused severe disruption to economic and financial activity worldwide.
We assess what happened to the aggregate U.S. stock market during this
period, including implications for both short and long-horizon investors.
Using the model of Maheu, McCurdy and Song (2012), we provide smoothed
estimates and out-of-sample forecasts associated with stock market dynamics during the pandemic.
We identify bull and bear market regimes including their bull correction and bear rally components,
demonstrate the model's performance in capturing periods of significant regime change,
and provide forecasts that improve risk management and investment decisions.
The paper concludes with out-of-sample forecasts of market states one year ahead.
\end{abstract}

\clearpage
\newpage


\section{Introduction}

This paper dates and forecasts bull and bear markets for the COVID-19
pandemic period
based on aggregate equity return data from 1885-2020.
Using the model of \cite{MMS:2012} applied to weekly data from 1885-2020,
we document where the market was and what has happened to equity markets in 2020.

There are several reasons for using the restricted 4-state Markov-Switching (MS) model of
\cite{MMS:2012}. First, unlike \emph{ex post} dating methods
\citep{Pagan-Sossounov:2003,Lunde-Timmermann:2004}, it treats the market
states as latent and provides probability estimates for future states and regimes.
This probability framework provides a full specification of the data generating process. As such, it
generates probability statements about the market dynamics which are essential for investment and risk management decisions.

Secondly, in contrast to simple specifications which focus on two
states of the market, our model allows for four states including bull corrections and bear rallies.
Conventional methods of partitioning regimes do not identify intra-regime
dynamics that can be very
important for forecasts and investment decisions.
For example, if a rally (positive sub-trend) starts during a bear regime,
what is the probability that it will continue and transition to a
bull market regime as opposed to falling back to the bear market state?
Will a bull market correction (negative sub-trend) continue to a bear
regime or recover to stay in the bull market regime?

Thirdly, higher-order moments of the state-specific distributions also provide useful
information -- for example, risk assessments associated with different states. They
combine to provide the aggregate mixture distribution that governs the market
dynamics and the associated predictive mean and density forecasts
that are key for decisions.

Finally, the parameters and states are identified with economic restrictions
that are consistent with investors attitudes to market phases.
The data supports these economically motivated restrictions
compared to an unrestricted 4-state model.

We find that the market moved from
a bull state at the start of 2020 to a bull correction and quickly to a bear
market on February 26. This bear market dominated until June 3 when the
market transitioned to a bear rally and has remained in this phase. This is
in contrast to the dating methods used in the popular press (``New
  bull market in stocks could last three years and may produce another 30\%
  in gains, veteran strategist says'', Aug 31, 2020,
  https://www.marketwatch.com/). Those methods focus solely on price trends and
ignore risk. Our model classifies the recent market into the bear rally due
to the larger variance in returns. This elevated risk is not consistent with
past bull market phases that display lower variability.

Traditional 2-regime models using \emph{ex post} dating methods are unable to
distinguish between bear rally and bull market states.
Our approach does this, in part, due to probability estimates of risk differences across those states.
We show the dramatic impact that the COVID-19 pandemic has had on the return
distribution and risk measures. The model provides very accurate forecasts of
turning points and these would have been available in real-time to investors.

Given the benefits of a full probability model of stock market phases,
it is natural ask whether forecasts from this mixture-distribution model can
improve investment and risk management decisions. To this end, we define a pseudo
Sharpe ratio to characterize in-sample estimates of the state density
parameters. We then extend this measure to an out-of-sample predictive Sharpe ratio,
derived from the predictive density of returns which is sensitive to the forecasted market states.
This measure can be useful in assessing the risk and return of
entering the market.

Several market timing investments are explored. We show that simple timing
rules directing when to exit and enter the market lead to improved investment
decisions relative to a buy and hold strategy in 2020. These results are
robust to different timing strategies and are a result of the precise turning
points our model identifies and forecasts. Each of these market timing
strategies are in real time and would have been available to an investor
using the model for forecasts.

The paper concludes with long-horizon forecasts of state probabilities one year ahead.
If the effects of COVID-19 were to disappear today,
then our model predicts months, and not weeks, till the stock
market returns to normal times.

Our paper is organized as follows. Sections 2 and 3 briefly review the
structure and estimation of the \cite{MMS:2012} bull and bear market model;
and Section 4 summarizes the data. Section 5 reports the results for the pre
and post COVID-19 periods. Notably, Section~5.3 provides
forecasts based on one-week-ahead predictive densities and
out-of-sample forecasts of future states and the associated risk and return measures.
Section~5.4 reports market timing strategies that
exploit those forecasts. Long-horizon forecasts are discussed in Section~5.5.
Section 6 provides robustness results,
comparing our model to competing models from several perspectives.
Section~7 concludes and an Appendix provides additional results and model
comparison details.

\section{\bf Model}

Define log-returns as $r_t$, $t=1,\dots,T$ and $r_{1:t-1}=\{r_1,\dots,r_{t-1}\}$.
Consider the following 4-state Markov-Switching (MS4) model from
\cite{MMS:2012} for returns,
\begin{eqnarray*}
  \label{eq:returns-4state}
  r_t | s_t & \sim & N(\mu_{s_t},\sigma_{s_t}^2) \\
p_{ij} & = &   p(s_t=j|s_{t-1}=i), \;\;\; i=1,...,4, \;\; j=1,...,4.  \\
 \end{eqnarray*}
in which $s_i, i=1,...,4,$ denote the latent states, parameterized as Normally distributed with mean
$\mu_{s_i}$ and variance $\sigma^2_{s_i}$, and $p_{i,j}$ denote the state transition probabilities.

The following restrictions and labels are imposed for identification purposes,
 \begin{eqnarray*}
  \label{eq:mu-restrictions}
\mu_1 & <& 0  \;\; \text{(bear state)}, \\   \nonumber
\mu_2 & > &0  \;\; \text{(bear rally state)},  \\  \nonumber
\mu_3 & <& 0  \;\; \text{(bull correction state)},  \\   \nonumber
\mu_4  & >& 0   \;\; \text{(bull state)}.    \\
\end{eqnarray*}
No restriction is imposed on $\{\sigma_1^2,\dots,\sigma_4^2\}$.
Note that there are four states but we refer to two distinct regimes by $B_t$
as
\begin{eqnarray*}
  \label{eq:mu-restrictions}
B_t =1  & \text{ if } & s_t=1 \text{ or } 2  \;\; \text{(bear regime)}, \\   \nonumber
B_t =2  & \text{ if }& s_t=3 \text{ or } 4  \;\; \text{(bull regime)}. \\
\end{eqnarray*}

The transition matrix takes the following form.
 \begin{eqnarray*}
\boldsymbol P & = &
\begin{pmatrix}
                       p_{11} & p_{12} & 0 & p_{14} \\
                       p_{21} & p_{22} & 0 & p_{24} \\
                       p_{31} & 0 & p_{33} & p_{34} \\
                       p_{41} & 0 & p_{43} & p_{44} \\
\end{pmatrix}
\end{eqnarray*}
This specification implies that bear states and bear rally states cannot move to
a bull correction and bull states and bull correction states cannot move to a bear rally
state. This is done to avoid confounding these states that display common mean
trends up or down in prices. For instance, it would be difficult to separate
states $1$ and $3$ which both display negative average growth. The restricted
transition matrix also means that the start of a new regime must be from a
bear state or a bull state.

To further enforce economic restrictions on this model, the following long-run
trends are imposed on the model. Solving for the stationary distribution
associated with $P$ we can compute the vector of unconditional state probabilities:
\begin{eqnarray}
  \label{eq:unconditionals-4state}
  \boldsymbol \pi = ( \boldsymbol A' \boldsymbol A)^{-1} \boldsymbol A' \boldsymbol e
\end{eqnarray}
where $ \boldsymbol A'=[ \boldsymbol P'- \boldsymbol I, \; \boldsymbol
\iota]$ and $ \boldsymbol e'=[0,0,0,0,1]$ and $ \boldsymbol \iota=[1,1,1,1]'$.
The long-run restrictions are
\begin{eqnarray}
  \label{eq:mean-restriction-bear}
  E[r_t|\text{bear regime}, B_t=1] & = & \frac{\pi_1}{\pi_1+\pi_2} \mu_1 + \frac{\pi_2}{\pi_1+\pi_2} \mu_2 <0 \\
  E[r_t|\text{bull regime}, B_t=2] & = & \frac{\pi_3}{\pi_3+\pi_4} \mu_3 + \frac{\pi_4}{\pi_3+\pi_4} \mu_4 >0.
 \label{eq:mean-restriction-bull}
\end{eqnarray}

\section{\bf Estimation}

We perform posterior
simulation with Gibbs sampling steps which reject any draws that violate the
parameter restrictions, coupled with the simulation smoother of \cite{Chib:1996} to sample
the latent state vector.
Estimation follows exactly from \cite{MMS:2012} with the same priors
employed in our analyses. For the MCMC output simulation,
consistent posterior moments or predictive density quantities can be computed
and are detailed in \cite{MMS:2012}. We collect $30,000$ posterior draws for
inference after dropping an initial $5,000$ draws for burn-in.

\section{\bf Data}

Daily equity capital gains for 1885 - 1927 are from \cite{Schwert:1990}. Equity data from
1928 - 2020 use the  S\&P500 index daily adjusted close reported by Yahoo Finance (\^GSPC
symbol). Risk-free return data are from the U.S. Department of the Treasury.
From these data, weekly continuously compounded returns, scaled by 100,
are obtained for 1885 - 2020. Weekly returns are computed using Wednesday data
and Thursday if Wednesday data is missing. The last observation is November
25, 2020.
In the following we refer to the equity index returns as the S\&P500 or the market.
A matching weekly realized variance measure $RV_t$, is computed as the sum of intra-week
daily squared returns.
Summary statistics for the weekly data are reported in Table~\ref{table_data_summary}.


\section{\bf Results}

Full sample parameter estimates for the four states are found in Table~\ref{tab:estimates}.
Those estimates include posterior means and 0.95 probability density
intervals for $\mu_i$ and
$\sigma_i$ associated with each of the four states.
Below those estimates, we report a pseudo Sharpe ratio, $\mu_{i}/\sigma_{i}$, for each state $i$ for $i=1,2,3,4$.
This measures the expected return adjusted for risk assuming a zero risk-free
rate for each state.
The parameter estimates are broadly similar to those in \cite{MMS:2012}.
The average return in the bear states, $-0.94$, is more negative than $-0.11$
in the bull correction phases of the bull regimes.
Analogously, the upward trend for returns in bull states ($0.52$) is stronger than $0.23$ in the bear rally phase of the bear regime.
Combining the return estimates with state volatilities, the Sharpe ratios also make sense, ranked from highest to lowest in the
bull, bear rally, bull correction and bear states respectively.

The posterior means of the state transition matrix $\boldsymbol P$
indicate persistence of bull and bear regimes in that
states within the bear regime ($s_t=1,2$) and those in the bull regime
($s_t=3,4$)
are likely to cluster together so that moves between
regimes will be infrequent. Interestingly, a bull correction
state is much more likely ($0.097$) to transition
back to the bull state than to move to a bear state
($0.013$); whereas, a bear rally state is more likely ($0.019$)
to transition to the bull state than it is to fall back to a bear state ($0.013$).
Furthermore, regime changes are almost always transitions from a bear rally or from a bull correction.
Directly jumping from a bull state to a bear state or vice versa is very rare.

Table~\ref{tab:uncond-probs} reports the unconditional probabilities
associated with the four states from which one can compute the
unconditional probabilities for the regimes. For example,
the bull market regime has a long-run probability of
$0.672$. Using the formulae in equation \ref{eq:mean-restriction-bear},
one can compute that the associated long-run mean of weekly returns in
bull regimes as $0.186$; whereas, that for bear regimes is $-0.069$.

\subsection{Before COVID-19}
\label{sec:before-covid19}

\begin{figure}
\caption{Estimates for 2019}
\includegraphics[width = 0.9\textwidth, height= 0.5\textheight]{\figurepath/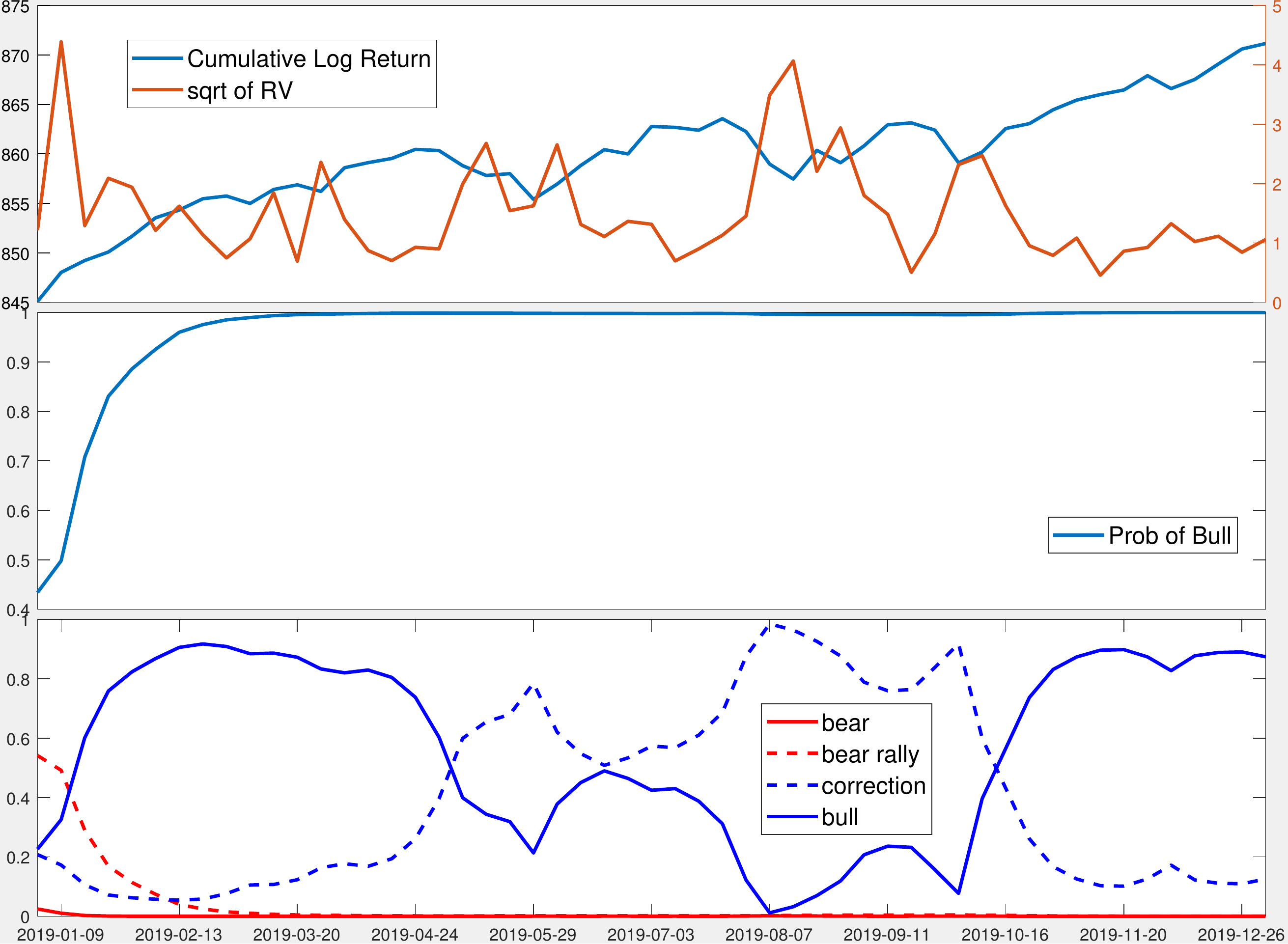}
\label{fig:2019}
\end{figure}

Figure~\ref{fig:2019} displays the cumulative log-return and realized volatility (top),
the probability of a bull regime $P(B_t=2\mid r_{1:T})$ (middle),
and the probabilities of the 4 individual states (bottom) during 2019, the year before the
COVID-19 pandemic.
Very early in 2019, the market moved from
a bear rally state into the bull state.
Throughout 2019 the model decisively identifies a bull regime with
fluctuations between the bull state and bull corrections.

\subsection{After COVID-19}

\begin{figure}
  \caption{Estimates for 2020}
\includegraphics[width = 0.9\textwidth, height= 0.5\textheight]{\figurepath/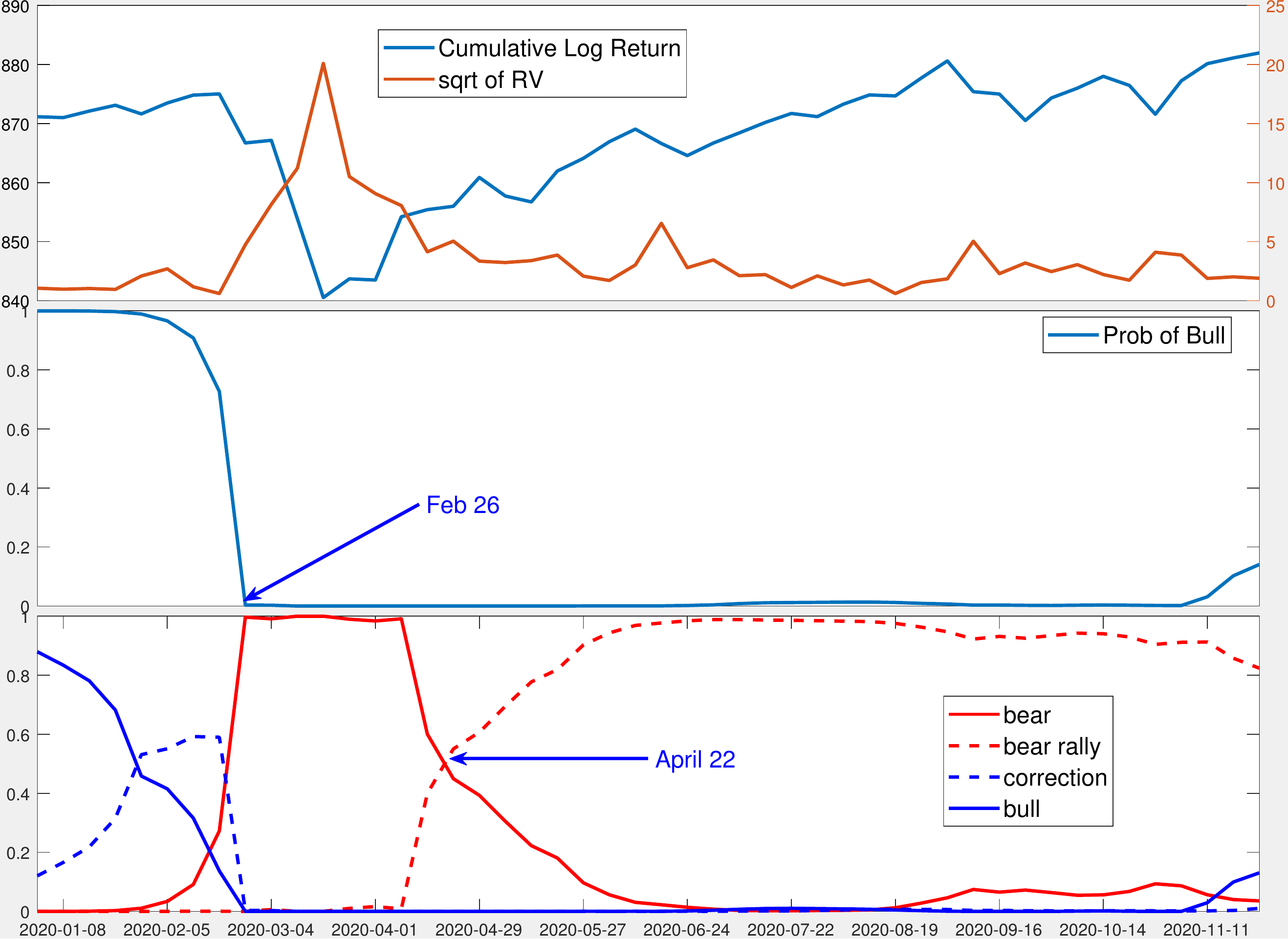}
\label{fig:2020}
\end{figure}

Figure~\ref{fig:2020} reports the same information as Figure~\ref{fig:2019} but for
the year 2020 during which COVID-19 erupted. The year began with strong
evidence of a bull market, although with the probability of a bull correction
building. The week of January 29 began a sequence of large negative drops in
the market leading to increasing evidence of a transition from a bull market to a bear
market. By February 26, the market had transitioned decisively from the
bull correction state to the bear market state.
By April the probability of the bear market state declined until
April 22 revealed a transition to a bear rally state.
At the end of our estimation window (November 25, 2020),
the probability of a bear rally state is still very high at $0.824$.
This is because, even with a visible upward trend in the index since April, the market
volatility is still too high and variable to be consistent with a typical bull regime.
This is evinced by the observation that the weekly realized volatility (red line in the top panel)
is variable, with spikes (for example, during September and October 2020), and higher on average in
comparison to its historical average value $1.94$ from Table~\ref{table_data_summary}.

\subsection{Forecasts}

Although one can date stock market cycles after the fact
with smoothed probability estimates,
it is much more difficult to forecast changes out-of-sample.
In this section, we report results for which the model has been estimated at each point $t$ using all
past data $1,\dots,t$ to produce a forecast of the state one week ahead $t+1$.

\begin{figure}
  \includegraphics[width = 0.9\textwidth, height= 0.5\textheight]{\figurepath/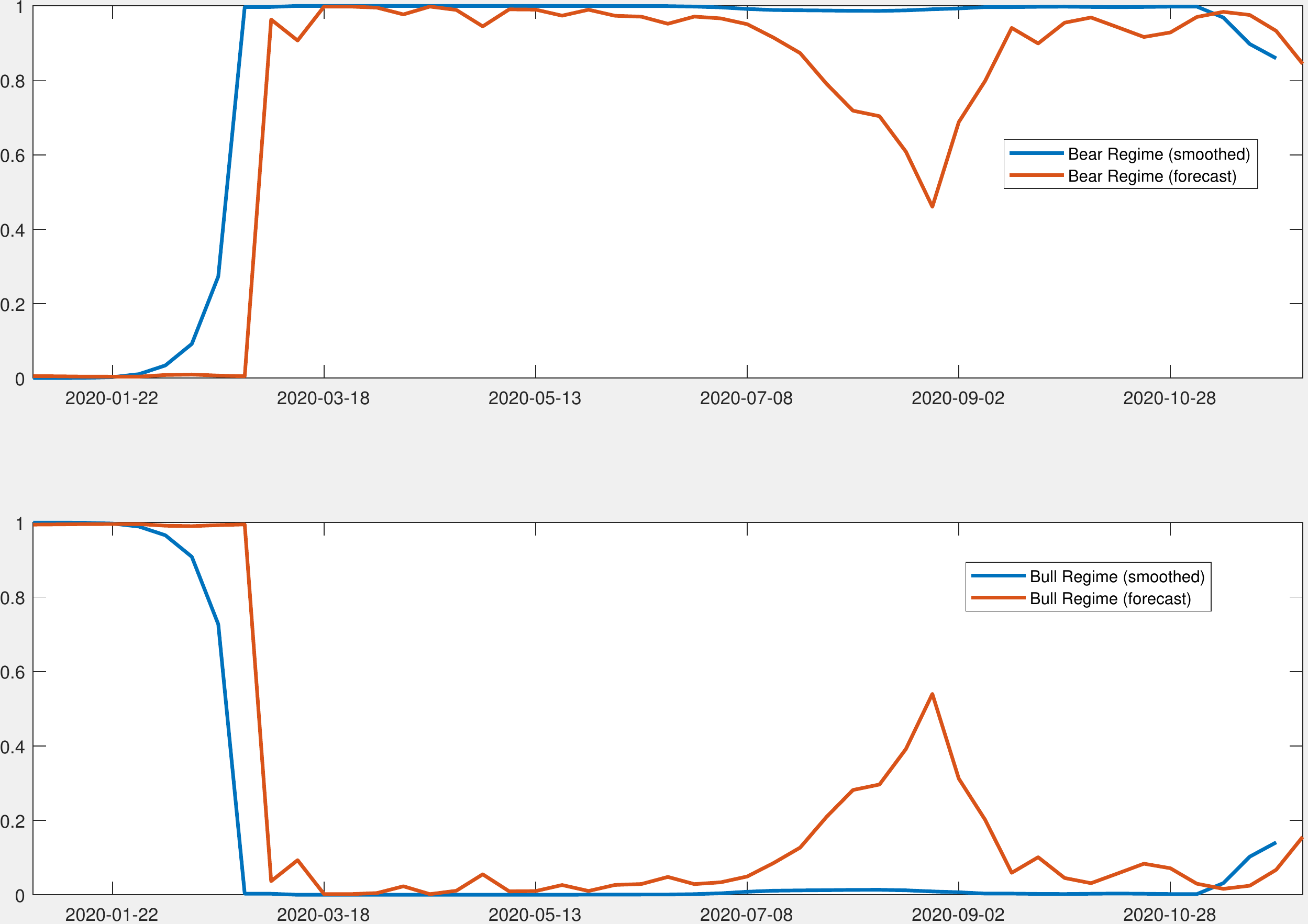}
\caption{Out-of-sample: One-week-ahead Regime Probability Forecasts}
  \label{fig:OOS-regimes}
\end{figure}

Figure~\ref{fig:OOS-regimes} uses these model forecasts of the state probabilities
one week ahead to generate out-of-sample regime forecasts.
This figure compares the out-of-sample forecast of the market regime probabilities with
the full-sample (smoothed) probability estimates. It shows a relatively accurate week-by-week
classification of regimes that was available in real-time to an investor
using the model to forecast stock market cycles.

One challenging period for the model forecasts was
late August to early September. From
July to September the index displayed a strong positive trend, and the model
forecasts allocated a
nontrivial probability to a bull regime. In hindsight, this episode
is precisely identified as a bear regime using the smoothed estimates.

\begin{figure}
  \includegraphics[width = 0.9\textwidth, height= 0.5\textheight]{\figurepath/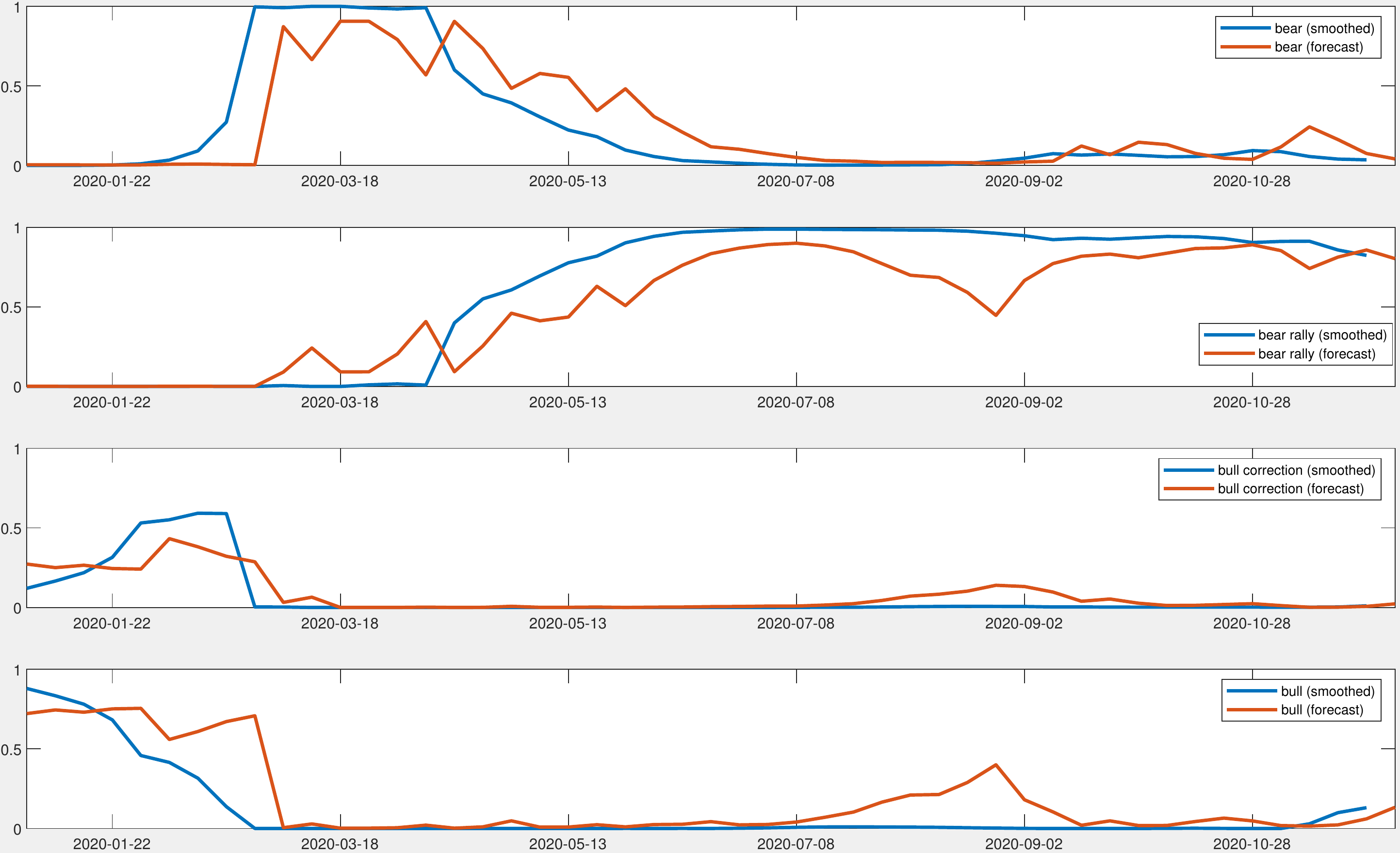}
\caption{Out-of-sample: One-week-ahead State Probability Forecasts}
  \label{fig:OOS-S}
\end{figure}

The breakdown of the regime forecasts into the constituent state forecasts
one week ahead in seen in Figure~\ref{fig:OOS-S}. Given this additional disaggregation,
there is more deviation between the state forecasts and smoothed estimates, but overall
there is a strong correspondence between the two. Note again that, with the exception of
the July to September period, the forecasts assign the highest probability to the bear rally state
rather than a bull state.

\begin{figure}
\includegraphics[width = 0.9\textwidth, height= 0.5\textheight]{\figurepath/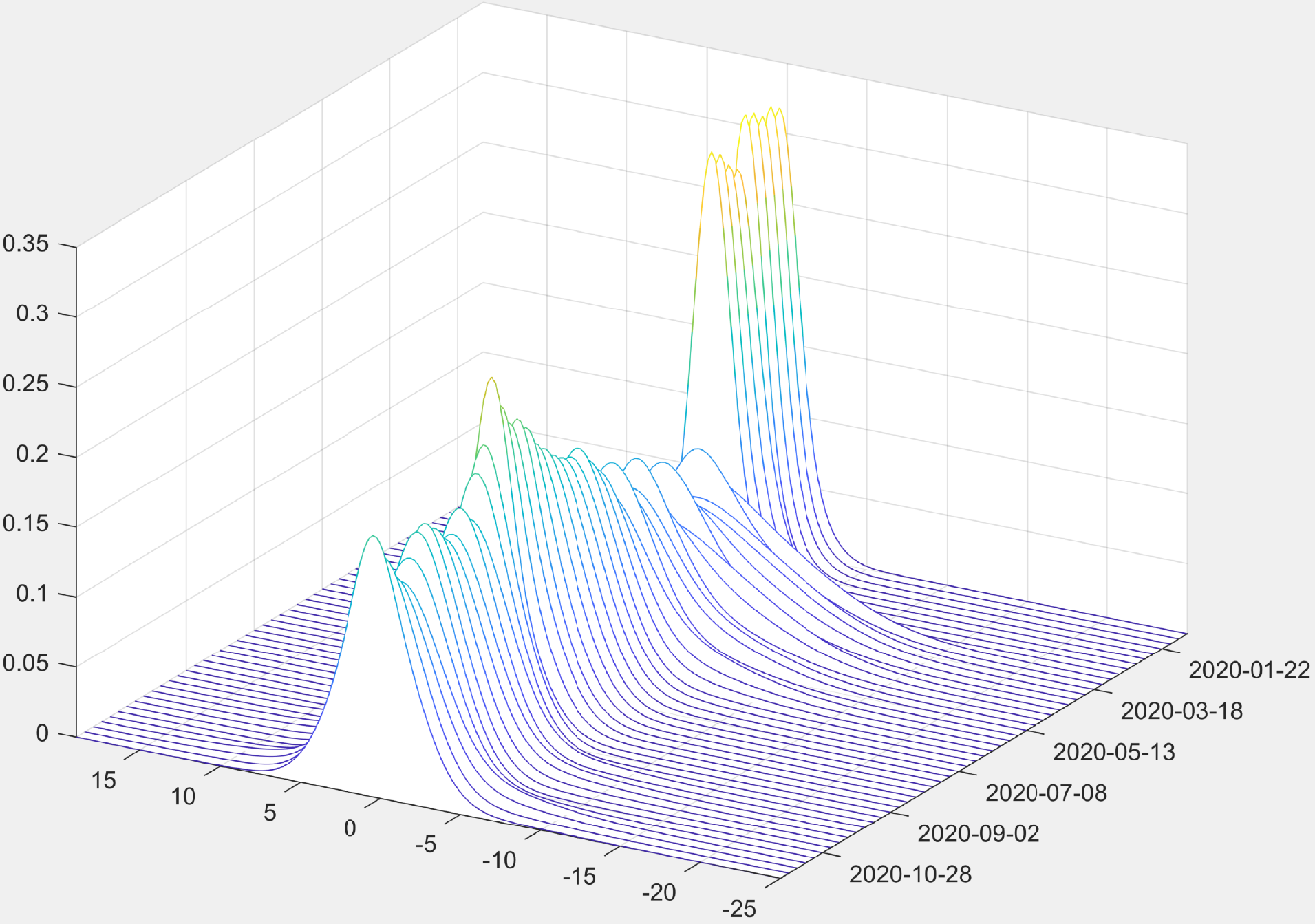}
\caption{Out-of-sample: One-week-ahead Predictive Densities }
\label{fig:pd1}

\medskip
\small{The X-axis is a grid of possible return values. The Y-axix is the time. The Z-axis is the probability density function values.}
\end{figure}

The COVID-19 period has had an important impact on several features of the return distribution.
For example, Figure~\ref{fig:pd1} displays the one-week-ahead predictive density generated by the
model. From the middle of March 2020 there was a dramatic impact that flattened the
return distribution for the rest of the year, along with more subtle changes in the
location. This implied a sharp increase in risk associated with holding the
S\&P500 portfolio.

\begin{figure}
  \caption{One-week-ahead Predictive Sharpe Ratios}
\includegraphics[width = 0.9\textwidth, height= 0.3\textheight]{\figurepath/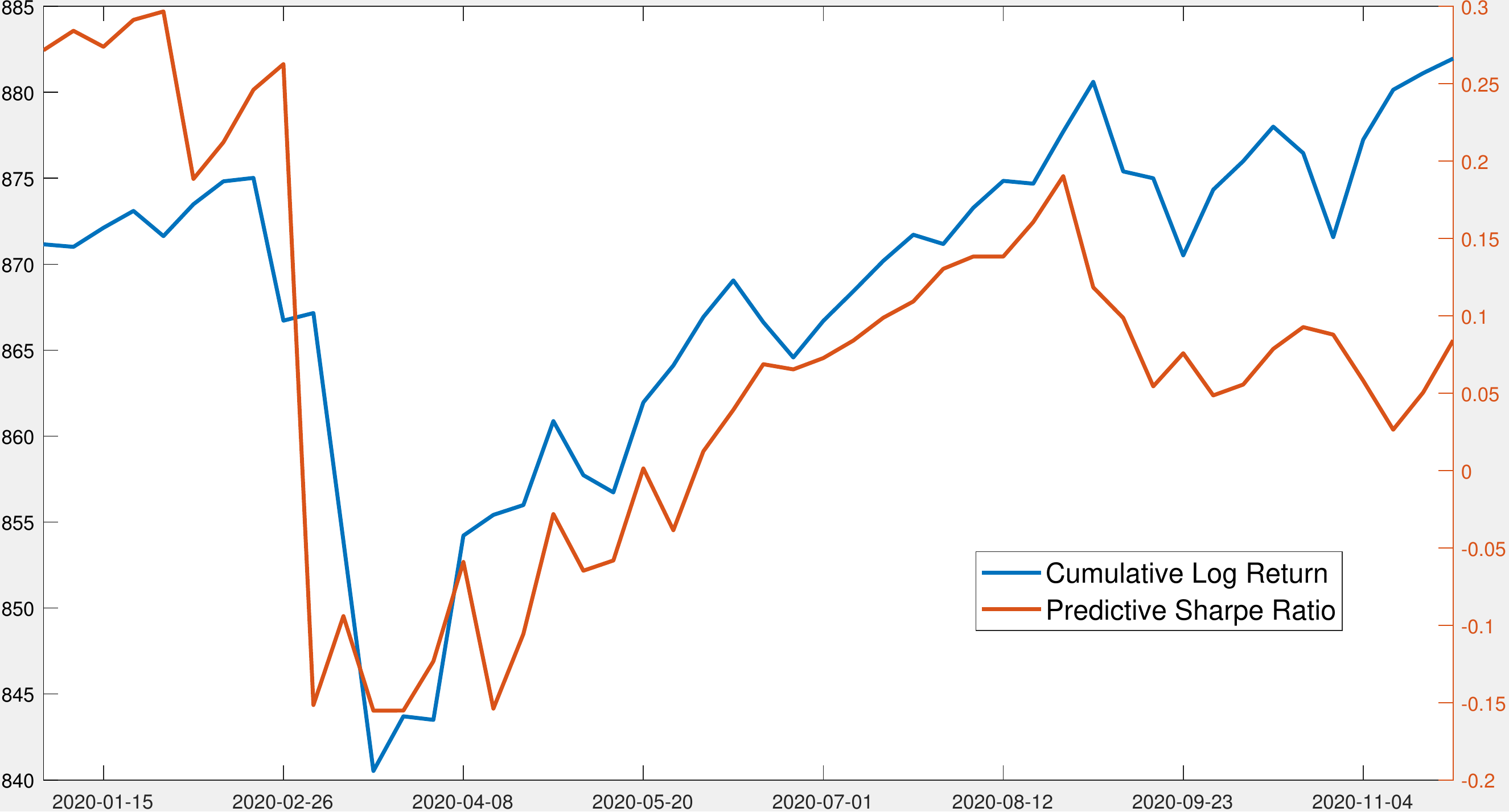}
\label{fig:2020_sharpe}

\medskip
\small{The Predictive Sharpe ratio is defined as the ratio of predictive mean and standard deviation,
$\frac{E(r_t\mid r_{1:t-1})}{\sqrt{\text{Var}(r_t\mid r_{1:t-1})}}$.}
\end{figure}

Figure~\ref{fig:2020_sharpe} illustrates the one-week-ahead predictive Sharpe
ratios defined as $\frac{E(r_t|r_{1:t-1})}{\sqrt{\text{Var}(r_t|r_{1`:t-1})}}$.
The flattened density of returns in March is accompanied by a sudden drop of the predictive Sharpe ratio.
This ratio becomes positive in June and continues to improve
over the summer months until a decline in August.
For the remainder of our sample, the predictive Sharpe ratio never does attain the values at the start of 2020
before COVID-19 struck.

\begin{figure}
\includegraphics[width = 0.9\textwidth, height= 0.4\textheight]{\figurepath/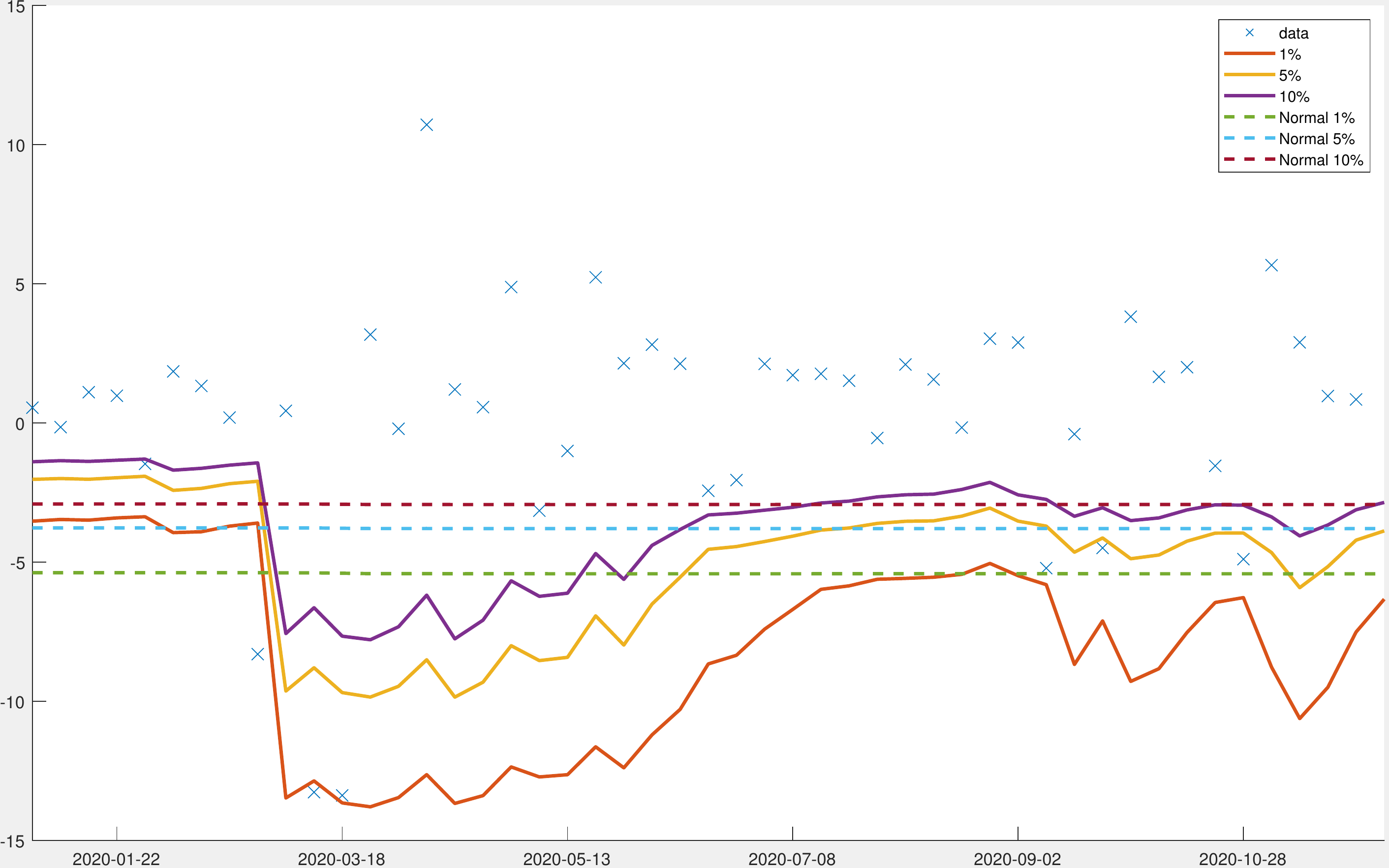}
\caption{Out-of-sample:  One-week-ahead Value-at-Risk Forecasts}
\label{fig:VaR}
\end{figure}

To illustrate another way, these changes in risk can be clearly seen from Value-at-Risk levels
estimated for our model and illustrated in Figure~\ref{fig:VaR}.
The dashed Value-at-Risk levels obtained from
an assumption that returns are normally distributed
would significantly understate risk as compared to those levels
implied by our model that incorporates a mixture of four state distributions.


\subsection{Market Timing}

In this section, we consider some simple market timing strategies that exploit
the forecasts from our model.
All of the investment strategies are based on the out-of-sample predictions of
states and regimes; and the simple maxim to buy low and sell high.
This can take several forms such as selling at the end of a bull market
and buying at the end of a bear market. However, our MS4 model provides much
more detailed and useful information. For example, the states that identify increasing prices are
state 4 (bull state) and the riskier state 2 (bear rally). Holding the market during periods for which forecasts
assign significant probability for these states could be fruitful.

In each case, the investor can buy the market and continue to hold the market; or sell and hold a risk-free asset.
No short selling is allowed. Here are the market timing strategies we consider.
\begin{enumerate}
\item Strategy B: buy or continue to hold the market when  \\
  $P(B_t=2|r_{1:t-1})> \tau_B$ and otherwise sell.
\item Strategy S: buy or continue to hold the market if \\ $P(s_t=2|r_{1:t-1})>\tau_S$ or
  $P(s_t=4|r_{1:t-1})>\tau_S$ and otherwise sell.
    \end{enumerate}
The first strategy B only uses the aggregate regime information associated with the probability forecasts for $B_t$. The second
strategy exploits the positive expected return in both the bear rally ($s_t=2$) and bull
states ($s_t=4$). We focus on using one cutoff value $\tau_S$ for both of those states but this could be
generalized and we present some evidence of this below.

Table~\ref{tab:inv} shows the investment results for 2020. All values are annualized.
The annualized return is 13.1\% with a Sharpe Ratio of 0.566 if the investor buys in the first week of 2020
and holds the position until the last week of our data sample (last Wednesday of November).
This compares with a hypothetical buy and hold return of 6.46\% if an
investor held the index for our entire sample from 1885.

As reported in Table~\ref{tab:inv}, using the market timing strategy B does not perform well in 2020 even with a range of alternative cutoff
values, $\tau_B$, for buying and selling. However, exploiting the additional information
about states provided by the MS4 model (as in strategy S), yields positive results. For example,
with $\tau_S=0.5$, the market timing strategy generates a 22\% annualized return with a Sharpe Ratio of 1.203.
This investment strategy performs significantly better than the buy and hold strategy.

\begin{figure}
  \caption{Market Timing Returns as a Function of Signal Threshold}
\includegraphics[width = 0.9\textwidth, height= 0.5\textheight]{\figurepath/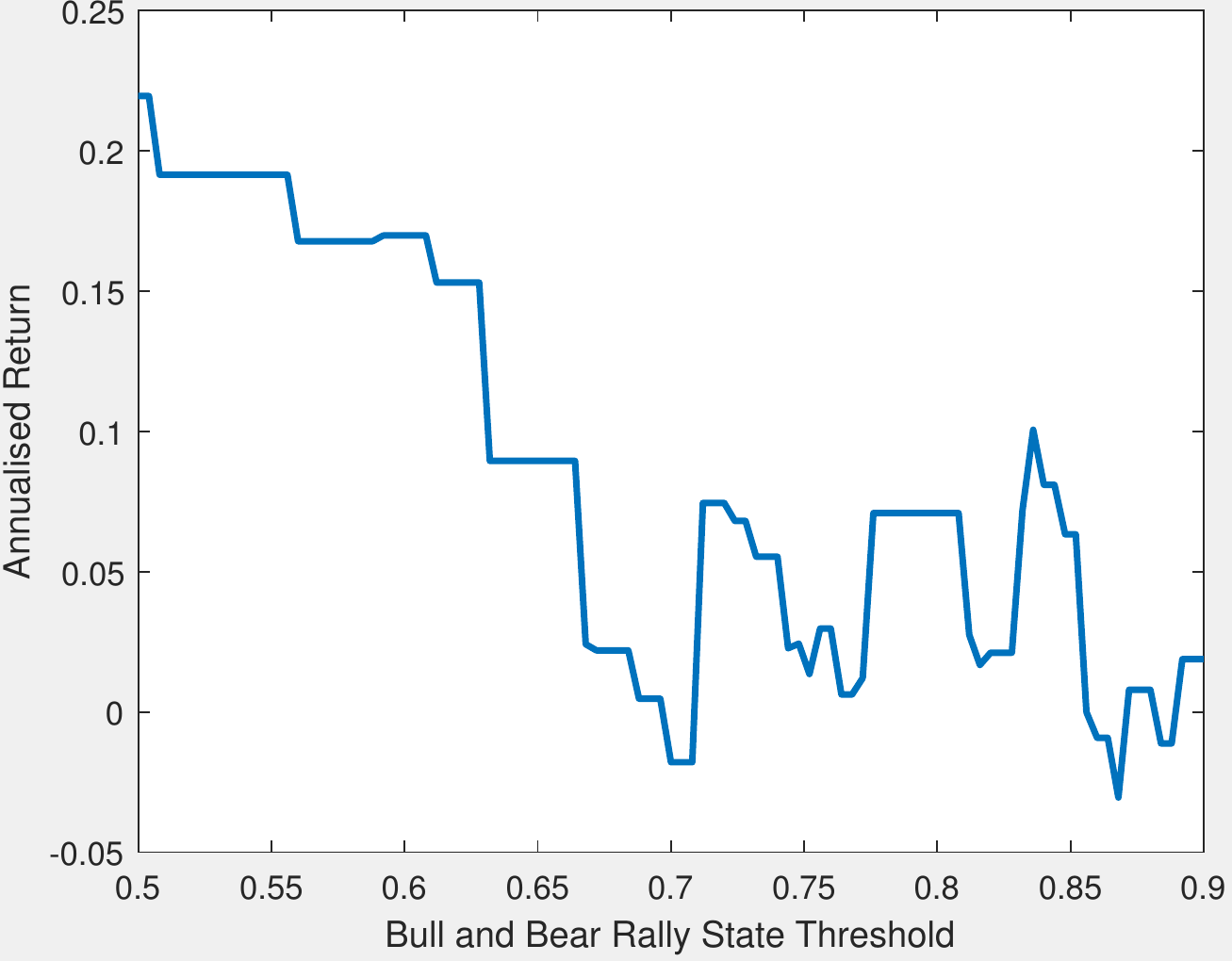}
\label{fig:threshold_taus}

\medskip
\small
The blue line is the return in 2020 till the end of the sample period as
a function of $\tau_S$ for the
investment strategy S: buy or continue to hold the market if $P(s_t=2|r_{1:t-1})>\tau_S$ or
  $P(s_t=4|r_{1:t-1})>\tau_S$ and otherwise sell.
\end{figure}

Figure~\ref{fig:threshold_taus} displays returns for strategy S as a function
of different values of $\tau_S$. For most values of $\tau_S \in (0.5,0.9)$ a
positive return is achieved with the best performance for values less than $0.65$.

\begin{figure}
 \caption{Market Timing Return as a Function of the Bull State Signal Threshold}
\includegraphics[width = 0.9\textwidth, height= 0.5\textheight]{\figurepath/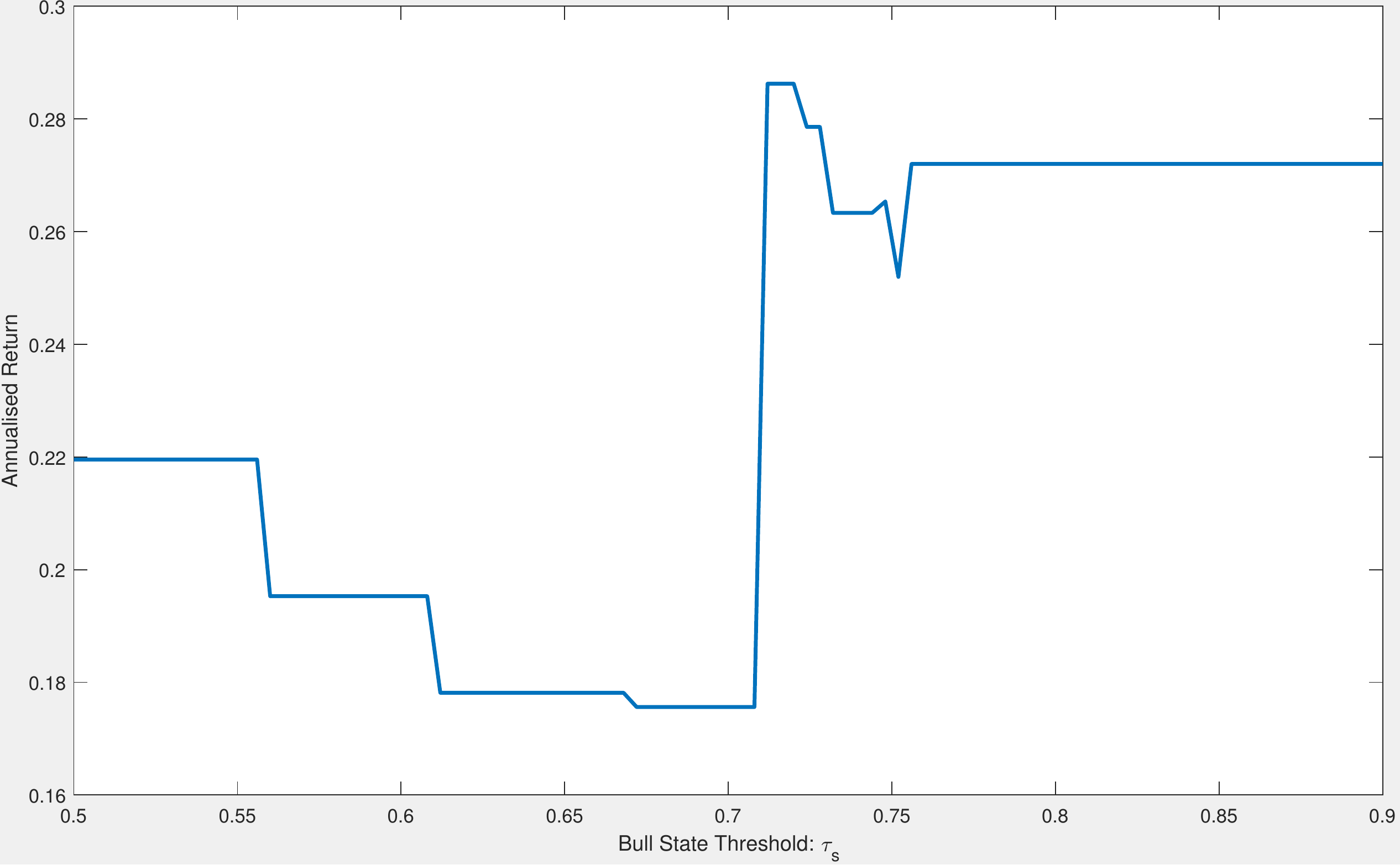}
\label{fig:threshold_bull}

\medskip
\small
The blue line is the return in 2020 till the end of the sample period as a
function of $\tau_S$
for investment strategy S:
buy or continue to hold the market if $P(s_t=2|r_{1:t-1})>0.5$ or
  $P(s_t=4|r_{1:t-1})>\tau_S$ and otherwise sell.
\end{figure}

Figure~\ref{fig:threshold_bull} shows the returns from strategy S while
relaxing the constraint of a common $\tau_S$ associated with both bear rally and bull states.
This figure fixes $\tau_S$ at $0.5$ for the bear rally state, that is, buying or continuing to hold the market
if $P(s_t=2|r_{1:t-1})>0.5$; while allowing $\tau_S$ to vary for the bull state, that is,
buy if $P(s_t=4|r_{1:t-1})>\tau_S$ and sell otherwise. The results show that it is possible
to achieve even larger gains by separating the thresholds in strategy S.
This is further evidence that the added value associated with using information inherent in the
state probabilities and predictive state densities for investment strategies is quite robust.

\subsection{Long-horizon Predictions}
\begin{figure}
  \caption{States Estimates for 2020 and One-year Forecasts}
\includegraphics[width = 0.9\textwidth, height= 0.5\textheight]{\figurepath/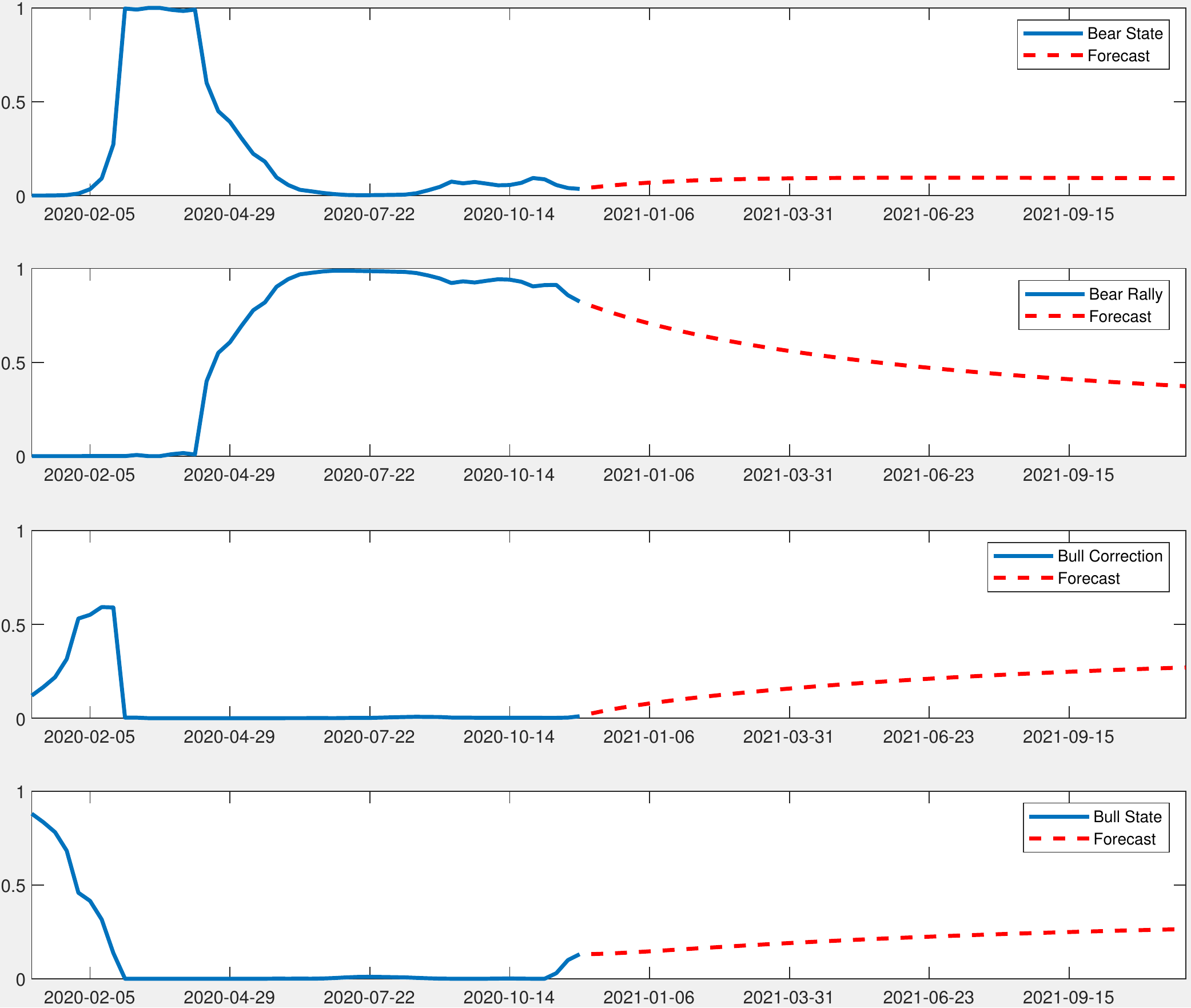}
\label{fig:2020_bbT52}
\end{figure}

\begin{figure}
  \caption{Regime Estimates for 2020 and One-Year Forecasts}
\includegraphics[width = 0.9\textwidth, height= 0.3\textheight]{\figurepath/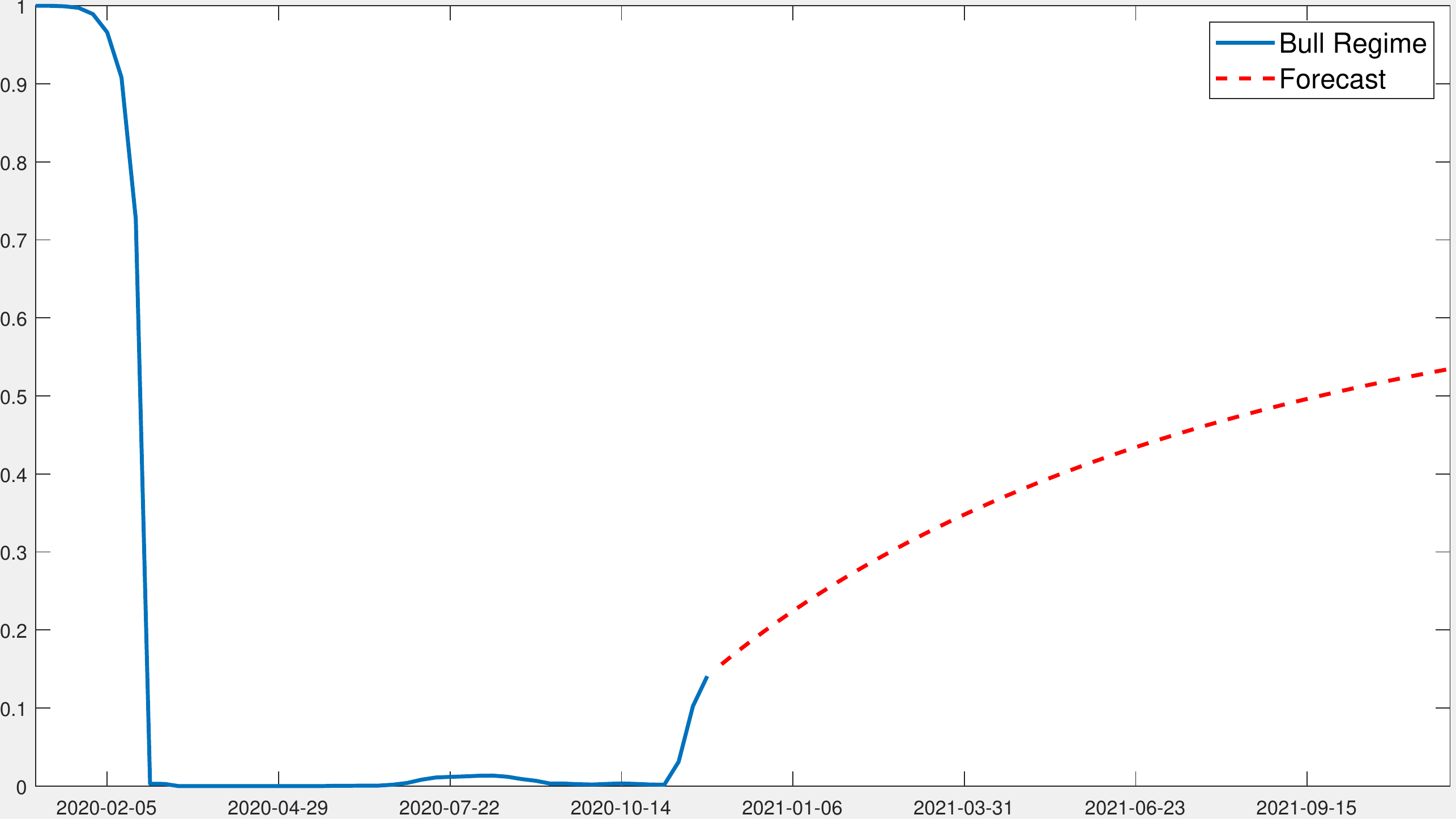}
\label{fig:2020_ST52}
\end{figure}

We have focused on one-week-ahead predictions from the model and how they might be
used for investment decisions. However, the model estimates have
implications for the long-run behaviour of stock market returns.
Being stationary, our MS4 model implies that any long-horizon predictions converge to the implied
stationary distribution. Figure~\ref{fig:2020_bbT52} and \ref{fig:2020_ST52} report one-year-ahead
probability forecasts for states and associated regimes looking forward from our most recent observations
at the end of November 2020. What is notable is that the transition
back to \emph{normal} times in the form of the long-run values of the states
is slow. Even if the effects of COVID-19 were to disappear today,
our model predicts months, and not weeks, until the stock market returns to normal times.

\section{\bf Robustness Results}

There are a number of alternative  models and comparisons we have
conducted. The details are collected in the Appendix. Here we highlight a few
of these results.
Table~\ref{tab:logPL} reports log-predictive likelihood values for the 2020
data for several alternative models. Included is our proposed 4-state model (MS4), an unrestricted version of MS4,
a MS4 model with Student-t innovations, a GARCH model and a MS2 model. None of these
specifications improve on our proposed MS4 model. In the Appendix, we show how these alternative
models date the turning points in 2020 and generally see a close
correspondence with the MS4 model. One notable exception is a 2-state Markov
switching model (MS2). The MS2 classifies the summer months as a bull state
while our MS4 identifies this period as a bear market rally. This could be considered as a drawback of
the 2-state model in that it does not allow for intra-regime dynamics.

\section{Conclusion}

This paper estimates and forecasts bull and bear markets during the COVID-19
pandemic. Using the model of \cite{MMS:2012}, we document where the market was
and what has happened to U.S. equity markets in 2020. We find that the market moved from
a bull state at the start of 2020 to a bull correction and quickly to a bear
market on February 26. This bear market dominated until April 22 when the
market transitioned to a bear rally, remaining in this phase until the current date.
We show the dramatic impact of the COVID-19 pandemic on the
forecasts of the return distribution and how effective market timing
strategies can exploit model forecasts. Long-horizon forecasts from the model
predict months, and not weeks, until the stock
market returns to normal times.

\section{Appendix}
\subsection{MS4 versus GARCH}
\begin{figure}
  \includegraphics[width = 0.9\textwidth, height= 0.3\textheight]{\figurepath/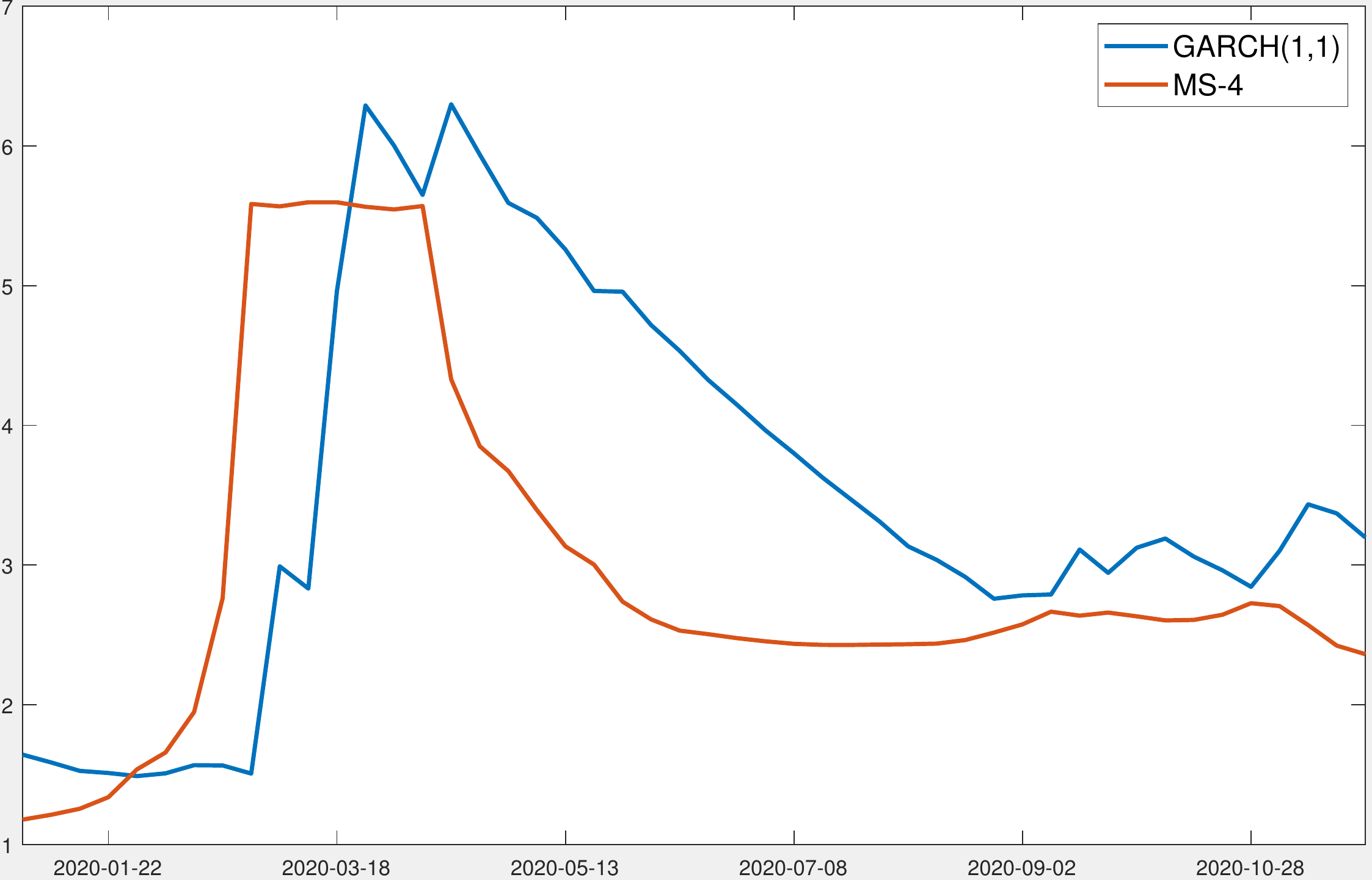}
\caption{Standard Deviations from GARCH(1,1) and MS4 in 2020.}
  \label{fig:arch11-ms4}
\end{figure}

\begin{figure}
\includegraphics[width = 0.9\textwidth, height= 0.5\textheight]{\figurepath/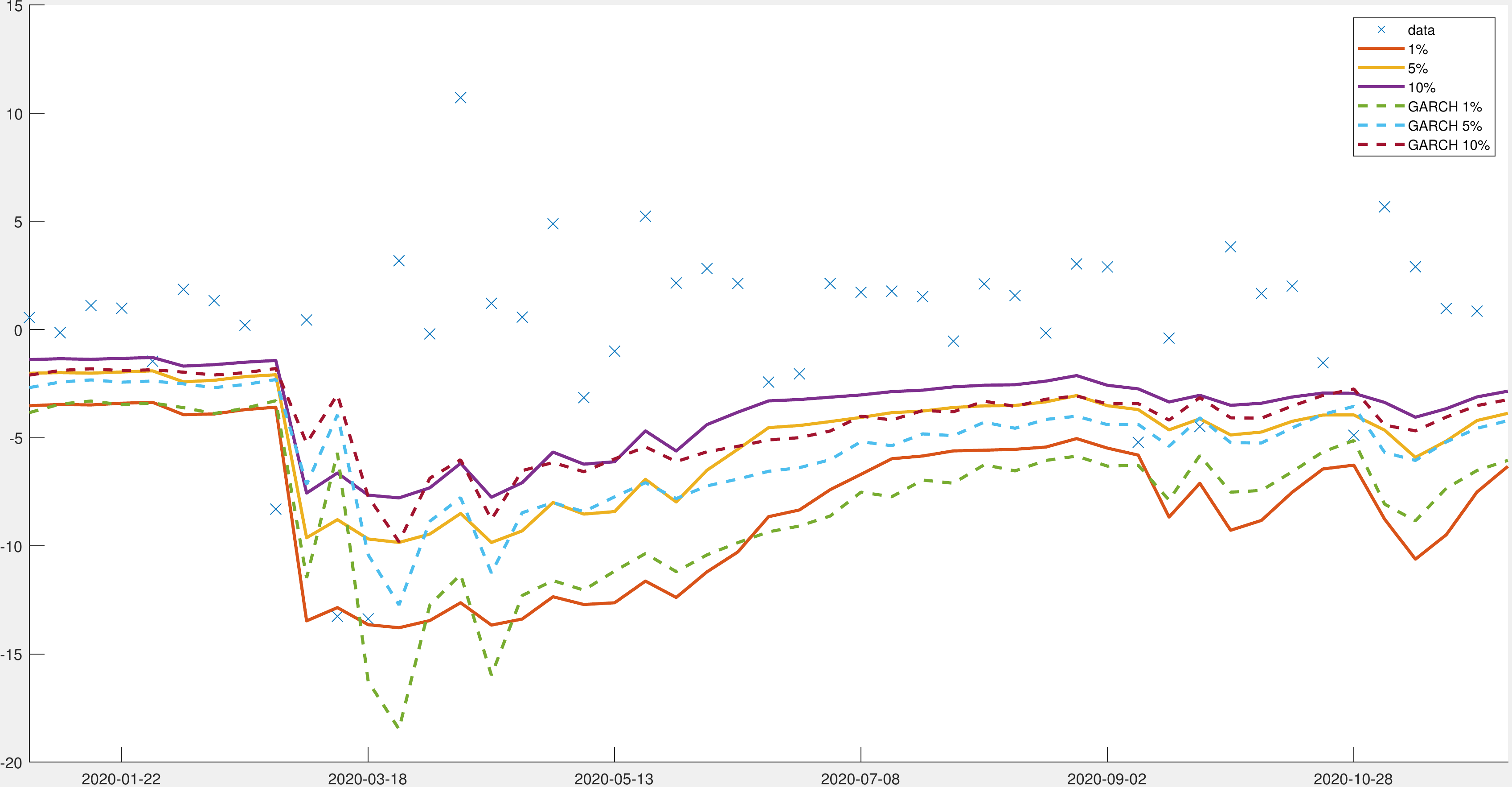}
\caption{Out-of-sample:  One-week-ahead Value-at-Risk Forecasts}
\label{fig:VaR_ms4_garch11}
\end{figure}

Figure~\ref{fig:arch11-ms4} shows the smoothed standard deviations from the MS4
versus a GARCH(1,1) model. The MS4 model picks up the volatility surge quickly
at the beginning of 2020 and does not have the long tapering-off period associated with the GARCH model.
The implications for Value-at-Risk are illustrated in Figure~\ref{fig:VaR_ms4_garch11}.
The quick adjustment of the Value-at-Risk estimates generated by the MS4
model match the high potential losses after March 2020.
This adjustment is substantially delayed and less flexible using a GARCH(1,1) model.
The latter is due to the single exponential decay parameter in the GARCH model which
invokes higher persistence and less flexible adjustment to shocks.

\subsection{MS4 versus MS2}
\begin{figure}
  \includegraphics[width = 0.9\textwidth, height= 0.4\textheight]{\figurepath/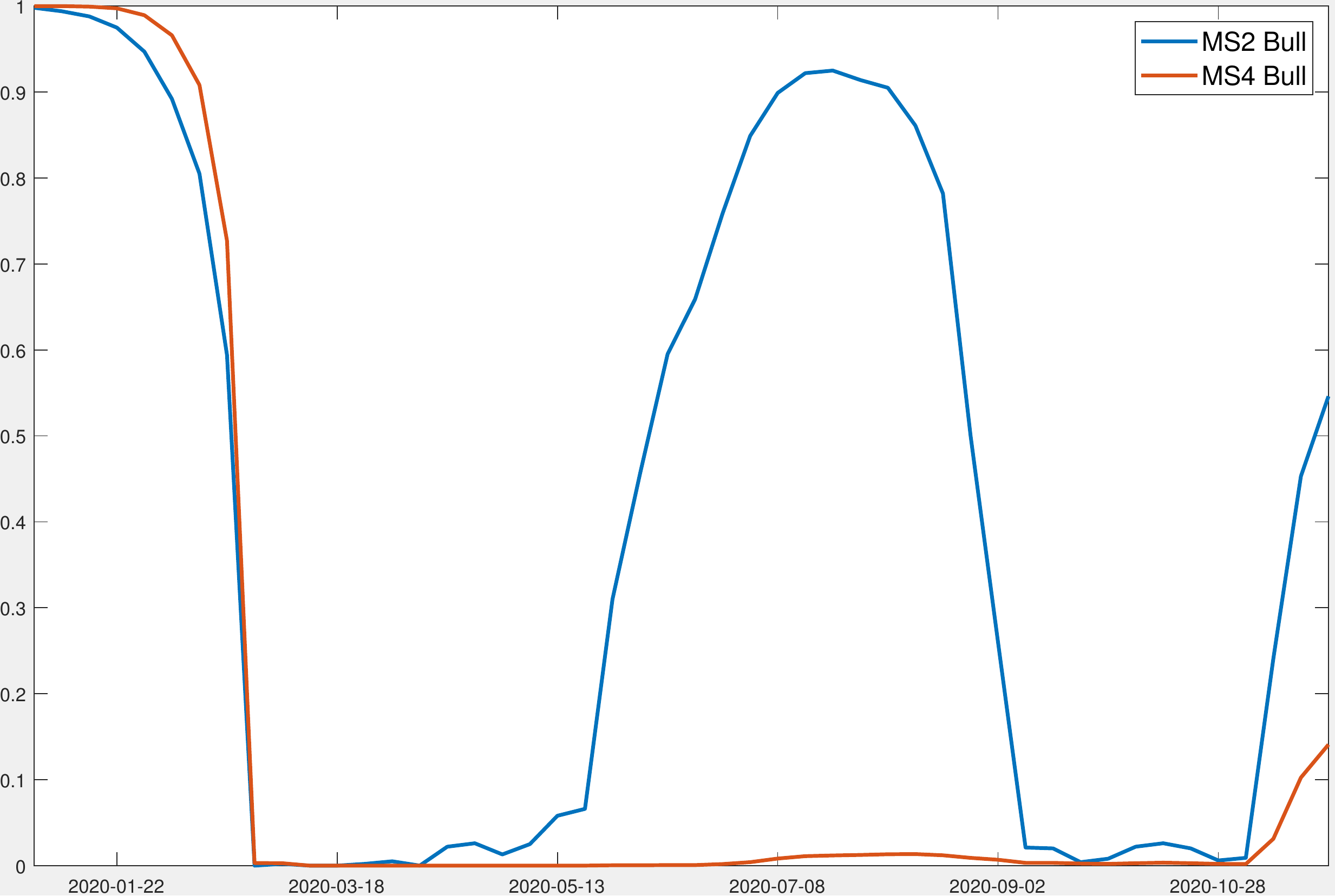}
\caption{Bull Regime Probability: MS2 vs MS4 Model}
  \label{fig:ms2-ms4}
\end{figure}

Figure~\ref{fig:ms2-ms4} shows the probability of a bull market regime inferred from
our 4-state model versus a simple 2-state model for the period 2019-2020.
There is one striking difference between these two models in the period from April
to September 2020. While the MS4 model classifies this period as a bear regime (in a bear rally state),
the MS2 model signals a bull regime. This difference is due to the 2-state model
not having the structure to incorporate intra-regime dynamics.
The September and October evidence from the 2-state model had the bull market probability
plunge to the bottom with small humps, while the 4-state model always
estimated that the bull regime had not been confirmed yet.

\subsection{MS4-t}
\begin{figure}
  \caption{State Probability Estimates: MS4t}
\includegraphics[width = 0.9\textwidth, height= 0.5\textheight]{\figurepath/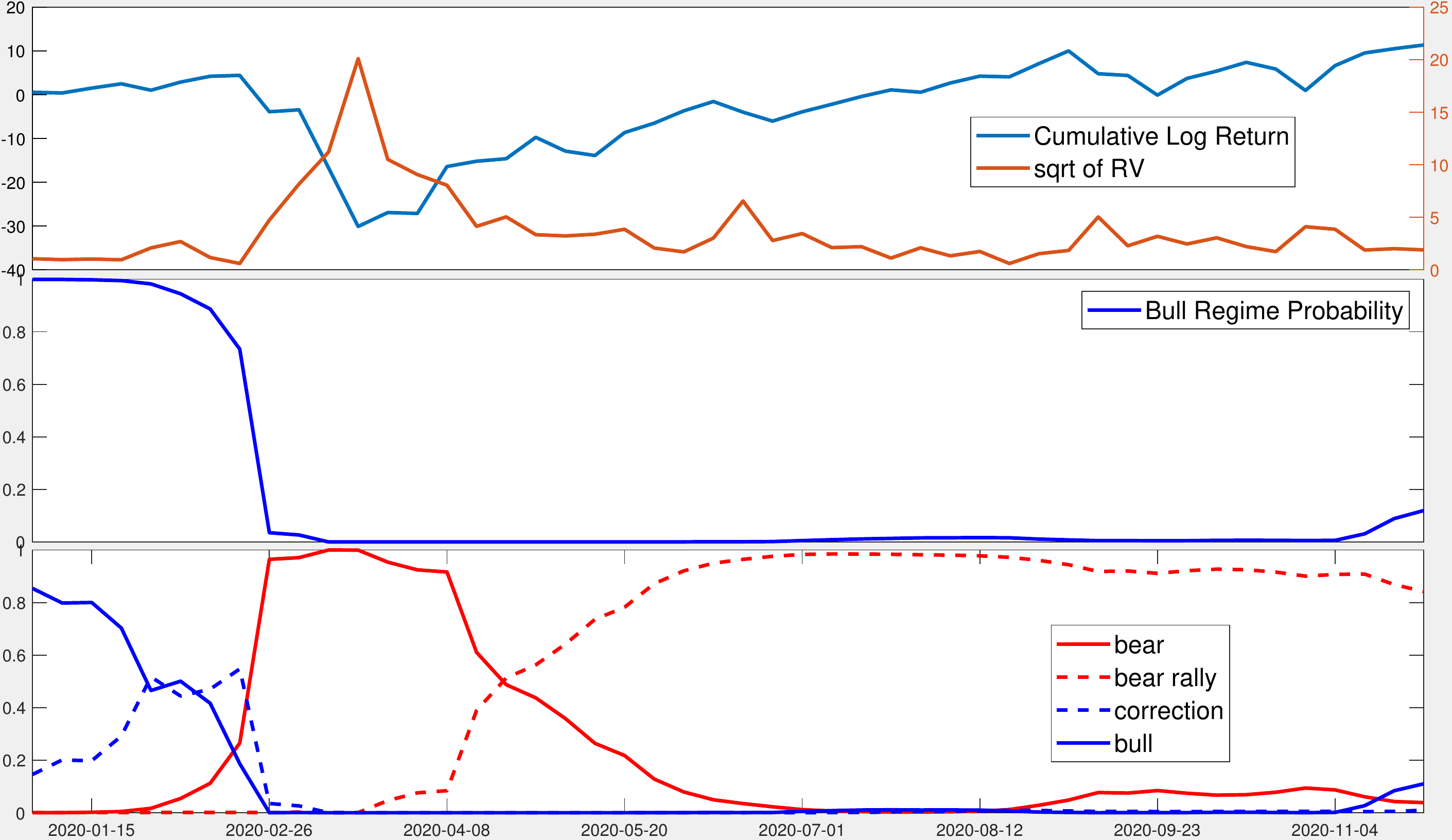}
\label{fig:2020t}
\end{figure}

Figure~\ref{fig:2020t} shows the
bull regime probability (middle) and the probability of each individual state (bottom)
associated with assuming a Student-t distribution for each state.
There is no qualitative difference from our proposed MS4 model in Figure~\ref{fig:2020}.

\subsection{Predictive Likelihood Comparisons}
Figure~\ref{fig:logPL} illustrates differences in the
cumulative predictive likelihood associated with alternative models for year 2020,
using the GARCH(1,1) model as the benchmark value.
The values at time $t$ are log predictive Bayes factor as
 $\log\left[\frac{p(r_{\text{2020 until }t}\mid r_{\text{before } 2020},\; \text{M})}
 {p(r_{\text{2020 until } t}\mid r_{\text{before } 2020},\; \text{GARCH(1,1)})}\right]$,
 where $M$ indicates various alternative models including the MS4.
 According to \cite{Kass-Raftery:1995}, a value $\frac{p(r\mid M_0)}{p(r\mid M_1)}$ that is larger than $5$
 indicates strong data evidence to support model $M_0$.
The 2020 data clearly favours the MS4 model against the GARCH(1,1).
In addition, having a Student-t distribution for each state does not provide any additional value.

\subsection{MS4 with unrestricted $ \boldsymbol P$}
\begin{figure}
  \caption{Cumulative Log Predictive Likelihoods}
\includegraphics[width = 0.9\textwidth, height= 0.5\textheight]{\figurepath/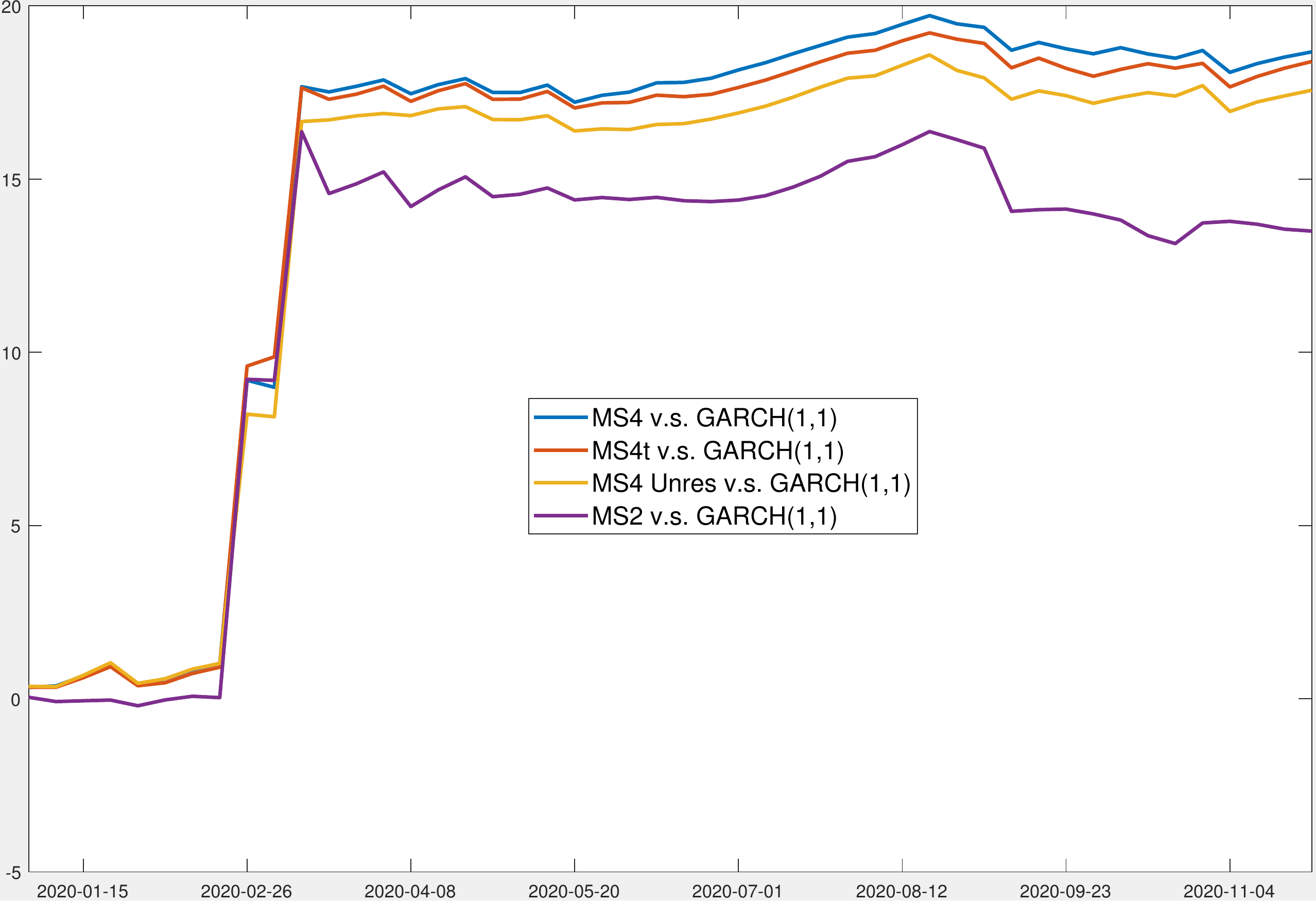}
\label{fig:logPL}
\end{figure}

\begin{figure}
  \caption{State Probability Estimates: MS4 with Unrestricted $ \boldsymbol P$}
\includegraphics[width = 0.9\textwidth, height= 0.5\textheight]{\figurepath/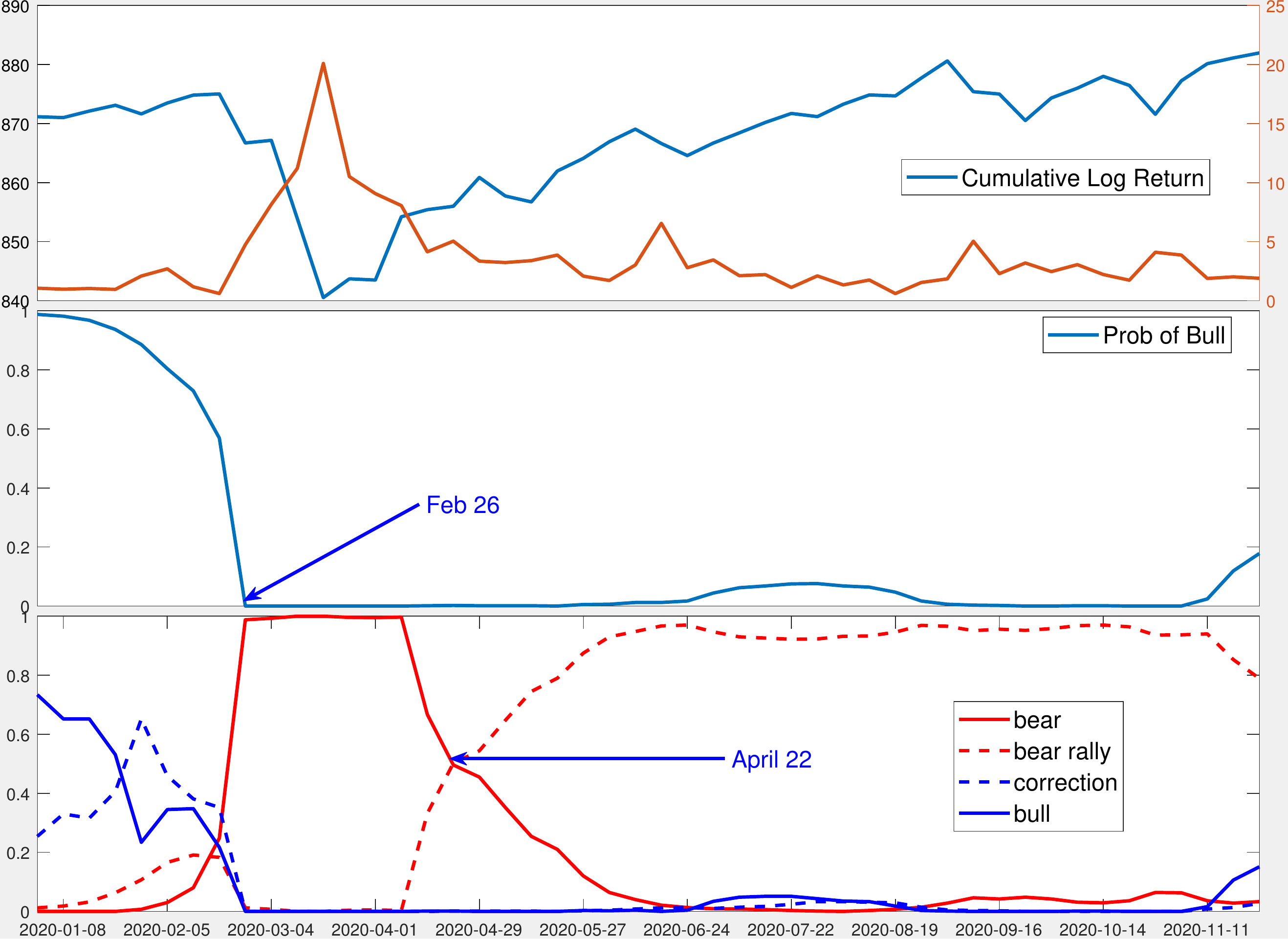}
\label{fig:2020_unres}
\end{figure}

\begin{figure}
  \caption{State Probability Estimates: MS4 with vs without $P$ Restrictions}
\includegraphics[width = 0.9\textwidth, height= 0.5\textheight]{\figurepath/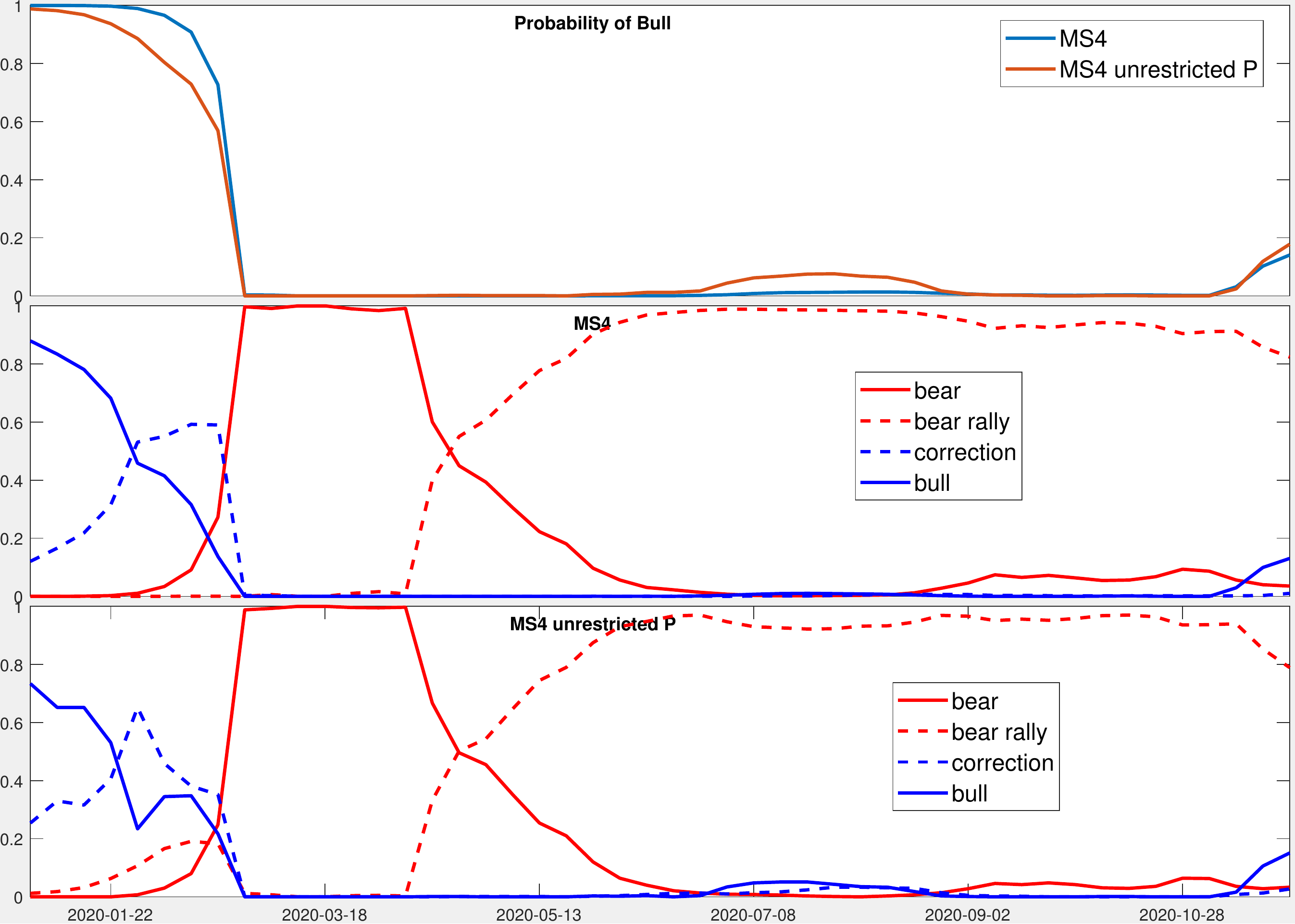}
\label{fig:2020_2ms4}
\end{figure}

Economic intuition suggests a zero restriction on $p_{13}, p_{23}, p_{32}$ and $p_{42}$. Without that restriction,
the 4-state model can still identify key dates of regime change in 2020 as shown in Figure~\ref{fig:2020_unres}.
However, a closer look at Figure~\ref{fig:2020_2ms4}, showing the bull regime and 4-state probabilities
from both MS4 and MS4 with unrestricted $ \boldsymbol P$, reveals that our proposed MS4 model, which has a restricted transition probability structure, has
less uncertainty between a bull market state and a bear rally state.
From the figure, during February and July 2020
the MS4 with unrestricted $ \boldsymbol P$ has higher bull state probability than the MS4,
because the unrestricted transition matrix has no structure to prevent
transition from the bull state to a bear rally state.
As a result, at the aggregate level, (top panel of Figure~\ref{fig:2020_2ms4}),
the MS4 has sharper identification of regimes than without the restriction on
$\boldsymbol  P$.
Further evidence can be seen from Figure~\ref{fig:logPL}, which reveals that the out-of-sample
performance of the unrestricted MS4 model is dominated by our proposed MS4 model.

Interestingly,
the MS4 model can be interpreted as a two-level hierarchical hidden Markov
model (\cite{fine1998hierarchical}),
in which each level has two states with simple restrictions.
The first restriction is on the vertical down probability
to incorporate the zero restrictions on the transition matrix.
The second restriction is the mean restriction for state-identification purposes.
These restrictions add value to bull and bear regime identification,
and proves the usefulness of a priori restrictions motivated by economic intuition.

\clearpage
\newpage

\bibliography{bb-update3}

\clearpage
\newpage
{\footnotesize\ctable[
   caption = {{\footnotesize Weekly Return Statistics}},
    label   = table_data_summary,
    pos     = htbp
]{lccccr}{
}{
  \FL
  & N & Mean &  Mean($RV^{.5}$) & Skewness & Excess Kurtosis  \ML
   &7064  &    0.125 &       1.938 &     -0.565 &       8.007\LL
}}

{\footnotesize\ctable[
    caption = {Posterior Estimates},
    label   = tab:estimates,
    pos     = htbp
]{rrc}
{
\tnote[]{}
}
{
  \FL
               &  mean  & 95\% DI \ML
bear    $\mu_1$    &  -0.94   & (-1.09,    -0.79) \NN
bear rally    $\mu_2$    &   0.23   & (0.14,     0.32 ) \NN
bull correction    $\mu_3$    &  -0.11   & (-0.21,    -0.02) \NN
bull    $\mu_4$    &   0.52   & (0.42,     0.64)  \NN
    $\sigma_1$ & 5.60    & (5.21,     6.03)  \NN
    $\sigma_2$ & 2.44    & (2.27,     2.61)   \NN
    $\sigma_3$ & 1.85    & (1.69,     2.04)   \NN
    $\sigma_4$ & 1.09    & (0.97,     1.21)   \LL
    $\mu_1/\sigma_1$ & -0.17 & (-0.20,    -0.14) \NN
    $\mu_2/\sigma_2$ &  0.10 & (0.06, 0.13)\NN
    $\mu_3/\sigma_3$ & -0.06 & (-0.12, -0.01) \NN
    $\mu_4/\sigma_4$ &  0.49 & (0.35, 0.65) \LL
}
\begin{eqnarray*}
  \label{eq:P-matrix-4state}
\text{Transition matrix} \;\;  \boldsymbol P=
\begin{pmatrix}
  0.906 &  0.092 &  0 &   0.002 \\
  0.013 &  0.968 &  0 &   0.019 \\
  0.013 &  0 &  0.891 &   0.097 \\
  0.001 &  0 &  0.122 &   0.876 \\
\end{pmatrix}
\end{eqnarray*}
}

{\footnotesize \ctable[
    caption = {Unconditional State Probabilities},
    label   = tab:uncond-probs,
    pos     = htbp,
    width = 80mm
]{rc}{
\tnote[]{}
}{
\FL
          &    mean  \\ \FL
bear $\pi_1$  &   0.084  \\
bear rally $\pi_2$  &   0.245  \\
bull correction $\pi_3$  &   0.356  \\
bull  $\pi_4$  &   0.316  \\ \FL
}}

{\footnotesize \ctable[
    caption = {Investment Returns},
    label   = tab:inv,
    pos     = htbp,
     width = 120mm
]{lrr}{
\tnote[]{The returns are annualized.}
\tnote[a]{Buy if $P(B_t=2\mid r_{1:t-1})>\tau_B$ and sell otherwise. }
\tnote[b]{Buy if $P(s_t=2\mid r_{1:t-1})>\tau_S$ or  $P(s_t=4\mid r_{1:t-1})>\tau_S$. Sell otherwise.}.
}{
\FL
                &    Return & Sharpe Ratio  \\ \FL
Strategy~B\tmark[a]: $\tau_B=0.5$ & -0.009 & -0.048 \\
Strategy~S\tmark[b]:  $\tau_S=0.5$ & 0.220 & 1.203\\
Buy-and-hold & 0.131 & 0.566  \\
              \FL
}}

{\footnotesize \ctable[
    caption = {Log-Predictive Likelihood in 2020},
    label   = tab:logPL,
    pos     = htbp,
    width = 100mm
]{ccccc}{
\tnote[]{}
}{
\FL
     MS4 &  MS4 Unrestricted & MS4t & GARCH & MS2 \\ \FL
-127.6 &-128.7& -127.9 &  -146.3  & -132.8\\ \FL
}}

%

\clearpage
\newpage

%
%


%
%

\begin{figure}
  \includegraphics[width = 0.9\textwidth, height= 0.4\textheight]{\figurepath/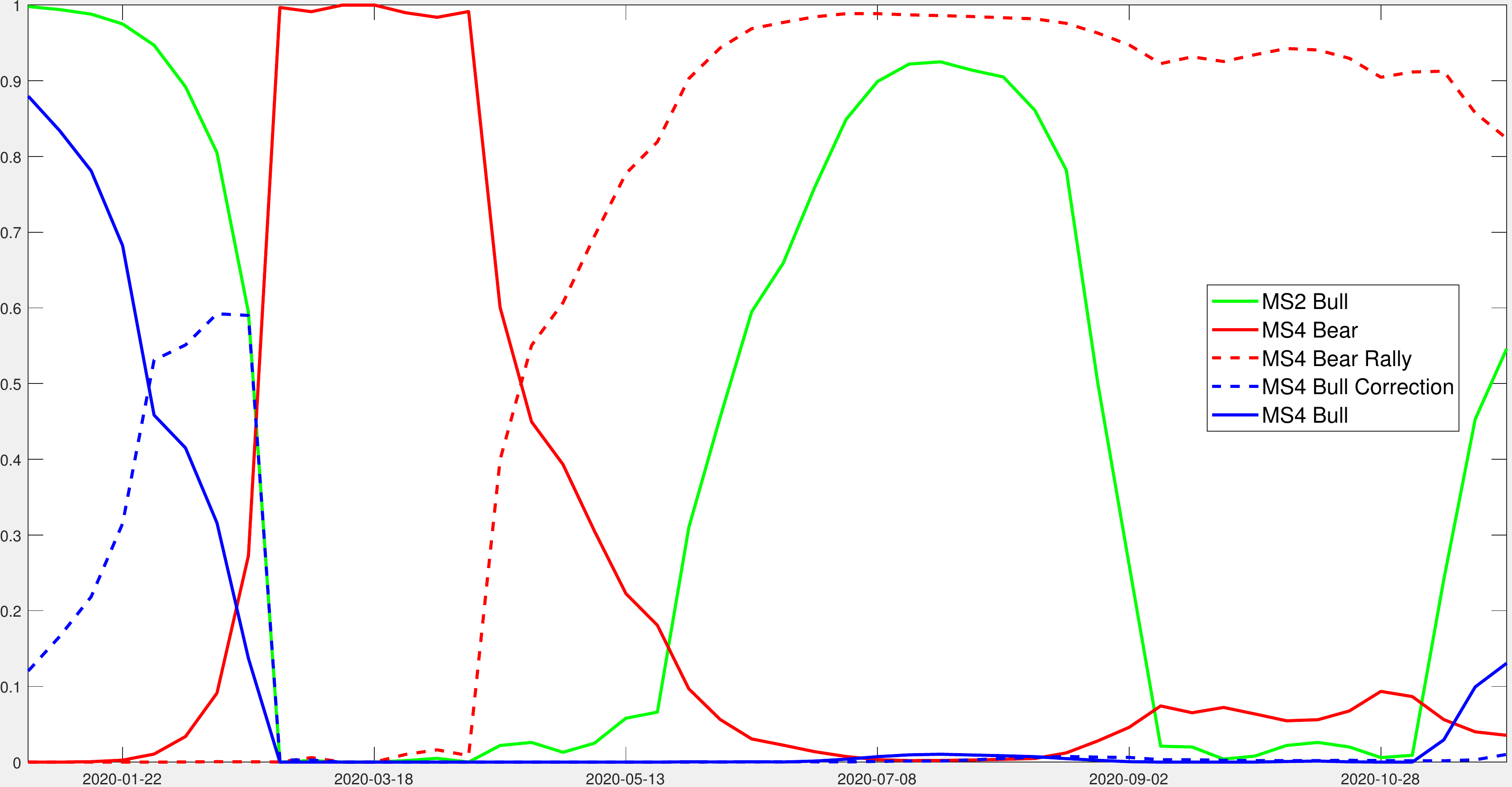}
\caption{Bull Regime Probability: MS2 \& State Probabilities: MS4}
  \label{fig:ms2-ms4-s2020}
\end{figure}

\clearpage
\newpage


\clearpage
\newpage




\end{document}